# The U(4) QCD Model at Finite Temperature

Dan-Na Liu[1] · Si-Zhao Huang[1] · Dian-Fu Wang[1]

**Abstract** Based on the U(4) strong interaction model, the behavior of the model at finite temperature is investigated. It is shown that, under high temperature, the dynamical breaking of gauge symmetry can be restored and the quark confinement can be melted away.

## 1 Introduction

At present, with the discovery of the 125 GeV Higgs boson in 2012 at the CERN LHC [1, 2], the Standard Model (SM) of particle physics stands triumphant, and all the data obtained from many experiments in particle physics are in agreement with this model. Despite its success, there are observational reasons to believe that the SM is not the whole story, such as neutrino masses, dark matter, and more fundamental ones such as Higgs fields and Higgs mechanism [3] its own problems.

The question is whether there is a natural way for us to introduce the Higgs fields to physical theories. Many authors have attempted to give the answer. The original work is an old idea due to Fairlie [4] and Ne'eman [5], of using SU(2/1) as a unification group for the Weinberg-Salam model and putting the Higgs fields in the adjoint along with the vector fields. Recently some authors have attempted to construct so called generalized Yang-Mills theories (GYMT) [6-10]. In [10], a maximally generalized Yang-Mills model (MGYMM) which contains, besides the vector part $V_\mu$, also an axial-vector part $A_\mu$, a scalar part $S$, a pseudo-scalar part $P$ and a tensor part $T_{\mu\nu}$ is constructed and the dynamical breaking of gauge symmetry in the model is also discussed. It is shown, in terms of Nambu-Jona-Lasinio (NJL) mechanism [11], that the gauge symmetry breaking can be realized dynamically in the maximally generalized Yang-Mills model. The combination of the maximally generalized Yang-Mills model and the NJL mechanism provides a way to overcome the difficulties related to the Higgs fields and the Higgs mechanism in the usual spontaneous symmetry breaking theory.

In our previous work [12], we have constructed a U(4) gauge invariant generalized Yang-Mills model about QCD, which contains both vector and pseudo-scalar gauge fields. Based on this model, we have obtained that, to the lowest-order approximation in $\hbar$, the numerical value of the vector gauge fields are equivalent mathematically to the pseudo-scalar fields. We have also obtained a U(4) strong interaction gauge potential solution which can be used to explain both the asymptotic behavior and quark confinement.

✉ Dian-Fu Wang
   wangdfu@dlmu.edu.cn
[1] Department of Physics, Dalian Maritime University, Dalian 116026, China

The purpose of the present paper is to investigate the behavior of the U(4) QCD model at finite temperature. The key point of our method is to solve the self-consistency equation for the order parameter by putting the invariant momentum cutoff at finite temperature. We will show that, the gauge symmetry broken dynamically can be restored at finite temperature. Meanwhile, we will also show that the quark confinement can be melted away under high temperature.

## 2 MGYMM and the U(4) QCD Model

Firstly, Let us briefly review the MGYMM we have proposed in [10]. In the usual Yang-Mills theory gauge invariance is assured through the demand that vector gauge transform as $\gamma_\mu V_\mu \to U(\gamma_\mu V_\mu)U^{-1} - i(\gamma_\mu \partial_\mu U)U^{-1}$. Recently, some authors [6-10] have again studied Yang-Mills theory and have constructed generalized Yang-Mills theories (GYMT) in which pseudo-scalar fields [6], axial-vector fields [7], scalar fields [9] and tenser fields [10] are considered to be also acceptable as gauge fields. Following these discussions, we have constructed a maximally generalized Yang-Mills model (MGYMM) [10].

The main idea of the MGYMM is as follows: Consider a Lagrangian which is invariant under a Lie group that has $N$ generators. Corresponding to each generator of the Lie group there is one gauge field, it does not matter whether vector fields or other fields. One can choose the first $N_V$ generators to be associated with an equal number of vector gauge fields and the last $N'$ generators to be associated with an equal number of the other fields. Naturally $N_V + N' = N$. By taking each of the generators and multiplying it by one of its associated gauge fields and summing them together, we construct a maximally generalized Dirac covariant derivative $D$ as

$$D = \gamma_\mu \partial_\mu - i\gamma_\mu V_\mu + \Phi, \tag{1}$$

where $\Phi$ denotes the generic gauge field which can be decomposed into irreducible Lorentz tensors

$$\Phi = S + i\gamma_5 P - i\gamma_\mu A_\mu \gamma_5 + \sigma_{\mu\nu} T_{\mu\nu}, \tag{2}$$

with $V_\mu = gV_\mu^a T^a$ the usual vector gauge field, $P = gP^b T^b$ a pseudo-scalar field, $S = gS^c T^c$ a scalar field, $A_\mu = gA_\mu^d T^d$ an axial-vector and $T_{\mu\nu} = gT_{\mu\nu}^e T^e$ a tensor field. The subscript $a$ varies from 1 to $N_V$, $b$ varies from $N_V + 1$ to $N_V + N_P$, and so on.

Define the transformation for the gauge fields as

$$-i\gamma_\mu V_\mu + \Phi \to U(-i\gamma_\mu V_\mu + \Phi)U^{-1} - i(\gamma_\mu \partial_\mu U)U^{-1}, \tag{3}$$

from which one can conclude that $D \to UDU^{-1}$. Just as in the usual Yang-Mills theories, when the covariant derivative $D$ acts on a certain field $\psi$, its gauge fields acquire certain coefficients called the charges of each gauge field with respect to $\psi$, namely

$$D\psi = (\gamma_\mu \partial_\mu - iQ_V \gamma_\mu V_\mu + Q_S S + iQ_P \gamma_5 P + Q_T \sigma_{\mu\nu} T_{\mu\nu})\psi. \tag{4}$$

In the MGYMM, the following Lagrangian is constructed based on the conditions that it contains only matter fields and covariant derivatives, and that it possess both Lorentz and gauge invariances:

$$L = -\bar{\psi} D\psi + \frac{1}{2g^2} \widetilde{Tr}\left(\frac{1}{8}(TrD^2)^2 - \frac{1}{2}TrD^4\right), \tag{5}$$

where the trace with the tilde is over the gauge (or the Lie group) matrices and the one without the tilde is over matrices of the spinorial representation of the Lorentz group. The expansion of (5) is

$$L = -\bar{\psi}\left(\gamma_\mu\partial_\mu - i\gamma_\mu V_\mu + Q_S S + iQ_P\gamma_5 P - iQ_A\gamma_\mu A_\mu\gamma_5 + Q_T\sigma_{\mu\nu}T_{\mu\nu}\right)\psi$$

$$-\frac{1}{2g^2}\widetilde{Tr}\left(\partial_\mu V_\nu - \partial_\nu V_\mu - i[V_\mu, V_\nu]\right)^2 - \frac{1}{4g^2}\widetilde{Tr}\left[Tr\left(\gamma_\mu\partial_\mu\Phi - i\{\gamma_\mu V_\mu, \Phi\}\right)^2\right]. \quad (6)$$

From the standard model we conclude that $Q_V = 1$.

Secondly, we will briefly review the U(4) QCD model which we have proposed in [12]. By considering the U(4) gauge invariant MGYMM with $\Phi = i\gamma_5 P$. The Dirac covariant derivative will acts on a basic quaternity which contains the three color components of a given quark flavour of the three family, called them $\psi_q^1, \psi_q^2, \psi_q^3$ plus a corresponding colorless state $\psi_q^0$ of the same multiplet. Here, we propose a hypothesis that the colorless state $\psi_q^0$ is the corresponding lepton state of the same generation, namely

$$\psi = \begin{pmatrix} \psi_q^1 \\ \psi_q^2 \\ \psi_q^3 \\ \psi_q^0 \end{pmatrix} \rightarrow \begin{pmatrix} u^1 \\ u^2 \\ u^3 \\ \nu_e \end{pmatrix}, \begin{pmatrix} d^1 \\ d^2 \\ d^3 \\ e \end{pmatrix}, \text{ and so on.} \quad (7)$$

The covariant derivative acting on (7) will be of the form $D = \gamma_\mu\partial_\mu - i\gamma_\mu V_\mu + i\gamma_5 Q_P P$ with $V_\mu = gV_\mu^a T^a$ $(a = 1, 2, \cdots, 8)$ and $P = gP^b T^b$ $(b = 0, 9, 10, \cdots, 15)$. $T^k$ are the generators of the Lie group-U(4) in the four dimensional representation. These are of the form $T^k = t^k/2$, where

$$t^a = \begin{pmatrix} & & & 0 \\ & \lambda^a & & 0 \\ & & & 0 \\ 0 & 0 & 0 & 0 \end{pmatrix}, \quad (8)$$

in which $\lambda^a$ $(a = 1 \cdots 8)$ are the Gell-Mann matrices, and

$$t^0 = \frac{1}{\sqrt{2}}\begin{pmatrix} 1 & 0 & 0 & 0 \\ 0 & 1 & 0 & 0 \\ 0 & 0 & 1 & 0 \\ 0 & 0 & 0 & 1 \end{pmatrix}, \quad t^9 = \begin{pmatrix} 0 & 0 & 0 & 1 \\ 0 & 0 & 0 & 0 \\ 0 & 0 & 0 & 0 \\ 1 & 0 & 0 & 0 \end{pmatrix}, \quad t^{10} = \begin{pmatrix} 0 & 0 & 0 & -i \\ 0 & 0 & 0 & 0 \\ 0 & 0 & 0 & 0 \\ i & 0 & 0 & 0 \end{pmatrix},$$

$$t^{11} = \begin{pmatrix} 0 & 0 & 0 & 0 \\ 0 & 0 & 0 & 1 \\ 0 & 0 & 0 & 0 \\ 0 & 1 & 0 & 0 \end{pmatrix}, \quad t^{12} = \begin{pmatrix} 0 & 0 & 0 & 0 \\ 0 & 0 & 0 & -i \\ 0 & 0 & 0 & 0 \\ 0 & i & 0 & 0 \end{pmatrix}, \quad t^{13} = \begin{pmatrix} 0 & 0 & 0 & 0 \\ 0 & 0 & 0 & 0 \\ 0 & 0 & 0 & 1 \\ 0 & 0 & 1 & 0 \end{pmatrix},$$

$$t^{14} = \begin{pmatrix} 0 & 0 & 0 & 0 \\ 0 & 0 & 0 & 0 \\ 0 & 0 & 0 & -i \\ 0 & 0 & i & 0 \end{pmatrix}, \quad t^{15} = \frac{1}{\sqrt{6}}\begin{pmatrix} 1 & 0 & 0 & 0 \\ 0 & 1 & 0 & 0 \\ 0 & 0 & 1 & 0 \\ 0 & 0 & 0 & -3 \end{pmatrix}. \quad (9)$$

With these objects we construct the corresponding Lagrangian of the U(4) QCD model which is invariant under the Lie group-U(4), namely

$$L = -\bar{\psi}\gamma_\mu \left(\partial_\mu - iV_\mu\right)\psi - i\bar{\psi}Q_P\gamma_5 P\psi$$

$$-\frac{1}{2g^2}\widetilde{Tr}\left(\partial_\mu V_\nu - \partial_\nu V_\mu - i[V_\mu, V_\nu]\right)^2 - \frac{1}{g^2}\widetilde{Tr}\left(\partial_\mu P - i[V_\mu, P]\right)^2. \quad (10)$$

In (10), since $\psi_q^0$ is the colorless state (the lepton state) with zero color charge, the constant $Q_P$ will be the form:

$$Q_P = \begin{pmatrix} Q_P & 0 & 0 & 0 \\ 0 & Q_P & 0 & 0 \\ 0 & 0 & Q_P & 0 \\ 0 & 0 & 0 & 0 \end{pmatrix}. \quad (11)$$

Substituting (11) into Euler equation, one can obtain the equations of motion for the vector gauge fields $V_\mu^a$, the fermion fields $\psi$ and the pseudo-scalar gauge fields $P_b$, respectively:

$$\gamma_\mu\left(\partial_\mu - igV_\mu^a T^a\right)\psi + iG_P\gamma_5 P^b T^b\psi = 0, \quad (12)$$

$$\left(\partial_\mu^2 - g^2 f V_\mu^a V_\mu^a\right)P^b - iG_P\bar{\psi}\gamma_5 T^b\psi = 0, \quad (13)$$

$$\left(\partial_\mu V_{\mu\nu}^a + gf^{abc}V_\mu^b F_{\mu\nu}^c\right) + g^2 f\left(P^b\right)^2 V_\nu^a - ig\bar{\psi}\gamma_\nu T^a\psi = 0, \quad (14)$$

in which $G_P = gQ_P$, $V_{\mu\nu}^a = \partial_\mu V_\nu^a - \partial_\nu V_\mu^a + gf^{abc}V_\mu^b V_\nu^c$ and $f = f^{abc}f^{abc}$. Multiplying the left and right-hand side of the equation (14) by $V_\nu^a$, and then taking vacuum expectation value of it, to the lowest-order approximation in $\hbar$, we obtain

$$\left\langle V_\mu^a V_\mu^a \right\rangle = \left\langle \left(P^b\right)^2 \right\rangle. \quad (15)$$

The equation (15) gives an important relation between the vector gauge fields and the pseudo-scalar gauge fields. As is well known, if the vacuum expectation value of the scalar fields is non-vanishing, the gauge symmetry will be broken down. Here one can choose the pseudo-scalar field $P^0$ to be associated with the unit generator $T^0$, we denote the vacuum expectation of the pseudo-scalar fields as

$$\left\langle P^b \right\rangle = \left\langle P^0 \right\rangle \neq 0. \quad (16)$$

From the equations of motion (14) we can easily obtain the mass of the vector gauge fields as

$$M_V = gf^{1/2}\left\langle P^0 \right\rangle. \quad (17)$$

By taking vacuum expectation value of the equation (13), to the lowest-order approximation in $\hbar$, we obtain the self-consistency equation as

$$M_P^2\left\langle P^0 \right\rangle = -iG_P\frac{\sqrt{2}}{4}\left\langle \bar{\psi}\gamma_5\psi \right\rangle, \quad (18)$$

where $M_P^2 = g^2 f\left\langle V_\mu^a V_\mu^a \right\rangle$. The equation (18) shows that the colorless pseudo-scalar field $P^0$ can be realized as the NJL type condensate of quark-antiquark pairs (mesons). Substituting the equation (15) into (18), one can rewrite the self-consistency equation as

$$g^2 f \langle P^0 \rangle^3 = -iG_P \frac{\sqrt{2}}{4} \langle \bar{\psi}\gamma_5\psi \rangle. \tag{19}$$

With an invariant momentum cut-off at $p^2 = \Lambda^2$, the self-consistency equation (19) is then changed to be

$$\begin{aligned} g^2 f \langle P^0 \rangle^3 &= -\frac{i\sqrt{2}}{4} G_P \langle \bar{\psi}\gamma_5\psi \rangle \\ &= \frac{i\sqrt{2}}{4} N_f G_P Tr(\gamma_5 S_F(0)) \\ &= -\frac{i}{2} N_f G_P^2 \langle P^0 \rangle \int \frac{d^4 p}{(2\pi)^4} \frac{1}{p^2 + G_P^2 \langle P^0 \rangle^2 / 8}, \end{aligned} \tag{20}$$

where $N_f$ is the number of hadronic constituents. Seeing from the equation (20) we can obtain the mass of the quark state as

$$M_\psi = \frac{\sqrt{2}}{4} G_P \langle P^0 \rangle. \tag{21}$$

Since the pseudo-scalar field $P$ describe the long-range collective effects of QCD, its short-wavelength components do not exist in reality. We can, therefore, consider only the tree approximation and ignore the loop approximation. Thus $\psi$, $P$ and $V_\mu$ can be reduced to $c$-number functions. Then in the ground state, by using the mean-field approximation [13], the equation (13) may be written as

$$\nabla^2 P^b - M_P^2 P^b = \frac{i\sqrt{2}}{4} G_P \langle \bar{\psi}\gamma_5\psi \rangle = -g^2 f \langle P^0 \rangle^3. \tag{22}$$

To solve the equation (22), it is convenient to divide the space into two regions: the inside region $R < R_0$, and the outside region $R > R_0$:

(1) Inside region $R < R_0$. In this area, the vacuum expectation value of the pseudo-scalar field $\langle P \rangle = gt^0 \langle P^0 \rangle / 2 = 0$. So from (17) and (21), we know that the quarks and the gauge fields are all massless, and the U(4) gauge symmetry is preserved. In this case, the equation (22) changes to be

$$\nabla^2 P^b = 0, \tag{23}$$

and the solution is:

$$P^b = -K_1 / R. \tag{24}$$

(2) Outside region $R > R_0$. In this area, the vacuum expectation value of the pseudo-scalar field $\langle P \rangle = (1/2) gt^0 \langle P^0 \rangle \neq 0$, where $\langle P^0 \rangle$ satisfy the self-consistency equation (20). In this case, the U(4) gauge symmetry, chiral symmetry and CP symmetry are broken down dynamically and the quarks and the gauge fields acquire a large mass. Meanwhile, the quark-antiquark condensate state, the colorless pseudo-scalar field $P^0$ exists. The solution of the equation (22) is given as

$$P^0 = -M_P^2 \langle P^0 \rangle \int \frac{dV}{R} g e^{-M_P R} = 4\pi f^{-1/2} M_P (M_P R + 1) e^{-M_P R},$$

$$P^b = 0, \quad b \neq 0, \tag{25}$$

in which $M_P = g f^{1/2} \langle P^0 \rangle$. Seeing from the equation (15), if we consider only the numerical value of the gauge fields $V_\mu^a$ and the pseudo-scalar fields $P^b$, we can assume that the gauge fields $V_\mu^a$ are equivalent mathematically to the pseudo-scalar fields $P^b$. One can write

$$V^a(R) \propto P^b(R). \tag{26}$$

Then we can combine (24), (25) and (26) into a uniform formula

$$V(R) \propto P^b(R) = -K_1/R + 4\pi f^{-1/2} M_P (M_P R + 1) e^{-M_P R}. \tag{27}$$

Inside the region of $R \leq R_h = R_0 + 1/M_P$, the factor $\exp(-M_P R)$ in formula (27) can be approximated to a constant, and formula (27) turns out to be

$$V(R) = -K_1/R + K_2 R, \tag{28}$$

where $K_1$ and $K_2$ are two arbitrary constants. As is well known the formula (28) is just the strongly interacting phenomenological gauge potential which has been used in discussing the energy levels of charmonium. Following from the formula (28) we can derive both the asymptotic freedom and quark confinement.

## 3 The U(4) QCD Model at Finite Temperature

In field theory, the Green function at finite temperature [14] can be written as

$$G_\beta(x_1 \ldots x_n) = \frac{1}{Z_\beta[J]} \frac{\delta^n Z^\beta[J]}{\delta J(x_1) \cdots \delta J(x_n)}, \tag{29}$$

where $\beta$ is proportional to the inverse of temperature $T$ and $Z^\beta[J]$ is the generation function at finite temperature

$$Z^\beta[J] = \int [d\varphi] \exp\left\{ \int_0^\beta d\tau \int d^3 x L_E[J] \right\}, \tag{30}$$

with $L_E[J]$ denotes the Lagrangian in Euclidean space.

The Green functions at finite temperature satisfy the same differential equations as the Green functions at zero-temperature except that they satisfy periodic (anti-periodic for the fermion case) boundary condition for an imaginary time $\tau$, and the momentum $p = (p_0, \vec{p})$ has to be replaced by

$$p = (\omega_n, \vec{p}), \tag{31}$$

in which $\omega_n = (2n+1)\pi/\beta$ (for fermions). Then, at finite temperature, by changing the integral of energy into sums, namely

$$\int dp_0 \rightarrow \frac{2\pi i}{\beta} \sum_{n=-\infty}^{+\infty}, \tag{32}$$

the self-consistency equation (20) changes to

$$2g^2 f G_P^{-2} N_f^{-1} \left\langle P^0(\beta) \right\rangle^2$$

$$= \frac{1}{\beta} \sum_{n=-\infty}^{+\infty} \int \frac{d^3 \vec{p}}{(2\pi)^3} \frac{1}{|\vec{p}|^2 + (2n+1)^2 \pi^2 / \beta^2 + G_P^2 \left\langle P^0(\beta) \right\rangle^2 / 8} = I_1 + I_2, \quad (33)$$

with

$$I_1 = \frac{1}{4\pi^2} \int_0^\infty \frac{|\vec{p}|^2 d|\vec{p}|}{\left(|\vec{p}|^2 + G_P^2 \left\langle P^0(\beta) \right\rangle^2 / 8\right)^{1/2}}, \quad (34)$$

$$I_2 = -\frac{1}{2\pi^2} \int_0^\infty \frac{|\vec{p}|^2 d|\vec{p}|}{\left(|\vec{p}|^2 + G_P^2 \left\langle P^0(\beta) \right\rangle^2 / 8\right)^{1/2} \left\{ \exp\left[\beta \left(|\vec{p}|^2 + G_P^2 \left\langle P^0(\beta) \right\rangle^2 / 8\right)^{1/2}\right] + 1 \right\}}. \quad (35)$$

The integration in (34) is divergent, by introducing an invariant covariant momentum cut-off $\Lambda$, one can make the integration finite. The result is

$$I_1 = \frac{1}{8\pi^2} \left[ \Lambda^2 - \frac{1}{8} G_P^2 \left\langle P^0 \right\rangle^2 \ln\left(\frac{\Lambda^2}{G_P^2 \left\langle P^0(\beta) \right\rangle^2 / 8} + 1\right) \right]. \quad (36)$$

Because we want to find the critical temperature $T_C$ at which the vacuum expectation value of the pseudo-scalar field $\left\langle P^0(\beta) \right\rangle$ tends to zero. So the integration in (35) can be calculated in the approximation $\left\langle P^0(\beta_C) \right\rangle = 0$. The result is

$$I_2 = -\frac{1}{2\pi^2} \int_0^\infty \frac{|\vec{p}| d|\vec{p}|}{1 + \exp(\beta_C |\vec{p}|)} = -\frac{1}{24\beta_C^2}. \quad (37)$$

Substituting (36) and (37) into (33) and taking the limit $\left\langle P^0(\beta_C) \right\rangle \to 0$, one gets

$$T_C = \frac{1}{\beta_C} = \frac{\sqrt{3}}{\pi} \Lambda. \quad (38)$$

Thus, we have found the critical temperature $T_C$ at which the mass of the vector gauge bosons and the pseudo-scalar gauge bosons $M_P(\beta_C) = M_V(\beta_C) = g f^{1/2} \left\langle P^0(\beta_C) \right\rangle$ tends to zero, which means that the dynamically broken $U(4)$ gauge symmetry can be restored at high temperature. Meanwhile the strongly interacting gauge potential (27) turns to be

$$V(R) \propto P^b(R)$$

$$= -K_1/R + 4\pi f^{-1/2} M_P(\beta_C)(M_P(\beta_C)R + 1) e^{-M_P(\beta_C)R} = -K_1/R, \quad (39)$$

which means that quark confinement at zero temperature can be melted away under high temperature.